\documentclass[prl,amssymb,amsmath,twocolumn,aps,showpacs,superscriptaddress,groupedaddress]{revtex4-1}

\usepackage{graphicx}
\usepackage{dcolumn}
\usepackage{bm}
\usepackage{hyperref}
\usepackage{subcaption}
\begin{document}

\title{General criterion for harmonicity}

\author{Karel Proesmans}
 \email{Karel.Proesmans@uhasselt.be}
  \affiliation{Hasselt University, B-3590 Diepenbeek, Belgium}
\author{Hans Vandebroek}
 \affiliation{Hasselt University, B-3590 Diepenbeek, Belgium}
\author{Christian Van den Broeck}

 \affiliation{Hasselt University, B-3590 Diepenbeek, Belgium, \\ Stellenbosch Institute of Advanced Studies, Matieland 7602, South Africa.}
 
\date{\today}

\begin{abstract}
Inspired by Kubo-Anderson Markov processes, we introduce a new class of transfer matrices whose largest eigenvalue is determined by a simple explicit algebraic equation. 
Applications include the free energy calculation for various equilibrium systems and a general criterion for perfect harmonicity, i.e., a free energy that is exactly quadratic in the external field. 
As an illustration, we construct a ``perfect spring", namely a polymer with non-Gaussian, exponentially distributed sub-units which nevertheless remains harmonic until it is fully stretched.
This surprising discovery is confirmed by Monte Carlo and Langevin simulations. \end{abstract}

\maketitle

The stretching of an (ideal) polymer provides one of the most beautiful illustrations of thermodynamics and equilibrium statistical physics.  A force is needed because the number of polymer configurations corresponding to a stretched state is (exponentially) smaller than that of a coiled state.
%
More generally, the extension $X$ versus force $F$ relation is obtained from  the derivative, at constant temperature $T$, of the free energy $\mathcal{G}$ \footnote{$\mathcal{G}=U-T S-FX$, with U the internal energy and S the entropy}:
\begin{equation}\label{fveeq}
X=-d\mathcal{G}/dF.
\end{equation}
Under isotropic conditions, one expects that the free energy $\mathcal{G}$ is even in $F$, hence quadratic for $F$ small. The extension $X$ is then harmonic, i.e., linear in the force $F$. For larger forces,  the polymer is expected to  stiffen as it approaches full extension. For example, for a freely-jointed chain in three dimensions, consisting of $N$ units of fixed length $b$, the fractional extension $x=X/Nb$ is given by  $x=   \mathcal{L}(b F\beta)$ with  $\mathcal{L}$  the  famous Langevin function $\mathcal{L}(y)=\coth(y)-1/y$ \cite{kuhn1942relationships} and $\beta=1/(k_BT)$, see also Fig.~\ref{langevinFig}.

The issue of elasticity is an important one in materials science. The width of the (harmonic) elastic regime, which can be estimated from the ratio of the  tensile strength over the Young's modulus, varies greatly from very small, for example for nanotubes, to very large for rubber and elastine  \cite{howard2001mechanics}. It is therefore natural to ask whether variations of a basic microscopic model, such as the freely-jointed chain, can lead to a predominant or even perfect harmonic response. The immediate answer appears to be yes: the Rouse model \cite{rouse1953theory} supposes bonds that are perfectly harmonic and hence so is the entire chain. But the assumption of perfectly harmonic bonds is unphysical as it would for example imply that both the bond and chain can be infinitely stretched. And deriving harmonicity from harmonicity is not exactly a great feat.
The surprising finding of this letter is the discovery of a simple random walk model for a polymer with non-harmonic bonds which is and remains perfectly harmonic up to full extension (and not beyond). 

Before proceeding to the more technical derivation, we comment on the route that led to this discovery and the additional results that were obtained. The statistical physics literature on polymers is huge, but exact results can only be derived for some very simple models such as the freely-jointed chain \cite{flory1969statistical,birshtein1966conformations}. One of the main tools to arrive at these results is the evaluation of the partition function via a transfer matrix method, essentially by identifying the largest eigenvalue.  Such transfer matrices have positive entries and are therefore reminiscent of Markov matrices, which describe the dynamics of Markov chains.  We introduce a special class of transfer matrices whose structure is inspired by a specific type of Markov process. We coin the name of ``Kubo-Anderson" transfer matrices in referral to two early papers (on line-width problems) in which such Markov processes have been introduced \cite{KA}.  The bonus is that the largest eigenvalue of such a transfer matrix is determined by a simple explicit algebraic equation. 
 Applications include  the free energy calculation for various equilibrium systems, including a simple model for polymer chains with persistence. Furthermore we derive, after an additional simplification, a general criterion for perfect harmonicity, i.e., a free energy that is exactly quadratic in the external field. We will discuss it here in the context of a simple random walk model for a polymer, but the results are equally valid for other systems, such as magnetic systems with exactly linear susceptibility. The application to simple polymer models leads to the discovery of the ``perfect spring", i.e., a random walk model for a polymer with non-Gaussian sub-units which nevertheless remains harmonic until it is fully stretched.

%

As is well known \cite{baxter2007exactly}, many problems in equilibrium statistical mechanics,  including the celebrated Onsager solution of the two-dimensional Ising model, can be formulated in terms of a transfer matrix. 
One supposes that the energy of the system can be written as a sum $E=\sum_{i=1}^N E_{\mathbf{\sigma}_{i+1},\mathbf{\sigma}_{i}}$, where $\mathbf{\sigma}_i$ represents the state of the ``i-th layer" and $E_{\mathbf{\sigma}_{i+1},\mathbf{\sigma}_{i}}$ is the interaction energy between layers $i+1$ and $i$. For notational simplicity, we consider periodic boundary conditions with layer $N+1$ identified with layer $1$. The central quantity is the partition function:
\begin{eqnarray}
   Z&=& \sum_{\{{\mathbf{\sigma}}\}}e^{-\beta\sum_{i=1}^NE_{\mathbf{\sigma}_{i+1},\mathbf{\sigma}_{i}}},
\end{eqnarray}
where the sum over $\{{\mathbf{\sigma}}\}$ runs over all configurational states of the system. 
The transfer matrix $\mathbf{T}$ is defined by its elements:
\begin{equation}\label{TE}
 \mathbf{T}_{{{\mathbf{\sigma}}',{\mathbf{\sigma}}}}=e^{-\beta E_{{\mathbf{\sigma}}',{{\mathbf{\sigma}}}}}.
\end{equation}
In the thermodynamic limit $N\rightarrow \infty$, the free energy $\mathcal{G}$ is obtained as:
\begin{eqnarray}
 \mathcal{G}=-\frac{\ln Z}{\beta}=-\frac{{\mathrm{Tr}\,\mathbf{T}^N} }{\beta}\sim -\frac{ N \ln \lambda}{\beta}, \label{PF}
\end{eqnarray}
where $\sim$ refers to  an equality  to dominant order in $N$, and  $\lambda$  is the largest  eigenvalue of the transfer matrix $\mathbf{T}$. 

A transfer matrix has positive elements.  The elements of a Markov matrix  represent probabilities, which are obviously also positive but obey in addition a normalization condition.
The relation between transfer matrices and Markov matrices has been noticed a long time ago, see, e.g., \cite{miller1961convexity,fisher1964magnetism}, and has been revisited more recently in the context of large deviations for conditioned Markov processes \cite{chetrite}.
In this letter, we introduce a new class of transfer matrices inspired by  ``Kubo-Anderson"  Markov processes. The latter have been used for a detailed analytic description  of a large variety of physical processes \cite{KA}. In the context of transfer matrices,  the bonus is a simple  explicit relation for its largest eigenvalue. 
Our starting point is the following Markov matrix:
  \begin{equation}
    \mathbf{A}_{\mathbf{\sigma}',\mathbf{\sigma}}=\left(1-q_{\mathbf{\sigma}}\right)\delta_{\mathbf{\sigma},\mathbf{\sigma}'}+q_{\mathbf{\sigma}} p_{\mathbf{\sigma}'}.\label{MarMat}
\end{equation}
$\mathbf{A}_{\mathbf{\sigma}',\mathbf{\sigma}}$ is the transition probability to go from state $\mathbf{\sigma}$ to $\mathbf{\sigma}'$. 
Its explicit form can be explained as follows. With a probability $1-q_{\mathbf{\sigma}} $ the system remains in its present state $\mathbf{\sigma}$. With probability $q_{\mathbf{\sigma}}$ a novel state is selected. This novel state is $\mathbf{\sigma}'$ with probability $p_{\mathbf{\sigma}'}$.
As is explained in the introduction with the stretching of a polymer, the addition of an external field allows to explore the regions of ``higher free energy".
We do something similar here by adding an extra Boltzmann factor to the transfer matrix,  with  $\epsilon_{\mathbf{\sigma}}$  representing the energy contribution due to such a field on layer $\mathbf{\sigma}$, leading finally to the following  ``Kubo-Anderson" transfer matrix:
\begin{eqnarray}
\mathbf{T}_{\mathbf{\sigma}',\mathbf{\sigma}}&=&e^{-\frac{\beta \left(\epsilon_{\mathbf{\sigma}}+\epsilon_{\mathbf{\sigma}'}\right)}{2}}\mathbf{A}_{\mathbf{\sigma}',\mathbf{\sigma}}.
\end{eqnarray}
The connection with the corresponding interaction energy $ E_{{\mathbf{\sigma}}',{\mathbf{\sigma}}}$ is  obtained by comparison with Eq.~(\ref{TE}). Conversely, this relation establishes the dependence of the quantities $q_{\mathbf{\sigma}} $ and $p_{\mathbf{\sigma}} $  on the energies $ E_{{\mathbf{\sigma}}',{\mathbf{\sigma}}}$ and the temperature.  
%

We now show how the largest eigenvalue (and corresponding eigenfunction) of  a ``Kubo-Anderson" transfer matrix $T$ can be obtained. From the eigenvalue equation
\begin{equation}
    \sum_{{\mathbf{\sigma}}} \mathbf{T}_{\mathbf{\sigma}',{\mathbf{\sigma}}}\mathbf{\phi}_{\mathbf{\sigma}}=\lambda \mathbf{\phi}_{\mathbf{\sigma}'},
\end{equation}
one finds that
\begin{equation}
e^{-\beta \epsilon_{\mathbf{\sigma}'}}(1-q_{\mathbf{\sigma}'})\mathbf{\phi}_{\mathbf{\sigma}'}+p_{\mathbf{\sigma}'}\sum_{\mathbf{\sigma}}e^{-\frac{\beta \left(\epsilon_{\mathbf{\sigma}}+\epsilon_{\mathbf{\sigma}'}\right)}{2}}q_{{\mathbf{\sigma}}}\mathbf{\phi}_{{\mathbf{\sigma}}}=\lambda \mathbf{\phi}_{\mathbf{\sigma}'}.\label{eveq}
\end{equation}
The largest eigenvalue $\lambda$ and its corresponding eigenfunction  $\mathbf{\phi}$ can be identified by invoking the Perron-Frobenius theorem: the eigenvalue $\lambda $ is unique and positive and all components of $\mathbf{\phi}$ have the same sign, and can thus be chosen to be positive. As a consequence $\sum_{{\mathbf{\sigma}}}\exp\left(-\beta \epsilon_{\mathbf{\sigma}}/2\right)q_{\mathbf{\sigma}}\mathbf{\phi}_{\mathbf{\sigma}}$ is also positive. Furthermore, since $\mathbf{\phi}$ is only determined up to a constant factor, we can assume the following normalisation:
\begin{equation}
\sum_{{\mathbf{\sigma}}}e^{-\frac{\beta \epsilon_{\mathbf{\sigma}}}{2}}q_{\mathbf{\sigma}}\mathbf{\phi}_{\mathbf{\sigma}}=1.\label{Nor}
\end{equation}
With this constraint,  we find from Eq.~(\ref{eveq}) the following explicit expression for the eigenvector $\mathbf{\phi} $:
\begin{equation}\label{eigenfunctioneq}
\mathbf{\phi}_{\mathbf{\sigma}}=\frac{p_{\mathbf{\sigma}}}{\lambda e^{\frac{\beta \epsilon_{\mathbf{\sigma}}}{2}}-e^{-\frac{\beta \epsilon_{\mathbf{\sigma}}}{2}}(1-q_{\mathbf{\sigma}})}.
\end{equation}
By substitution of this expression in  Eq.~(\ref{Nor}), one concludes that $\lambda$ is determined by:
\begin{equation}
\sum_{\mathbf{\sigma}} \frac{p_{\mathbf{\sigma}}q_{\mathbf{\sigma}}}{e^{\beta \epsilon_{\mathbf{\sigma}}}\lambda-(1-q_{\mathbf{\sigma}})}=1.\label{ceq}
\end{equation}
This simple explicit algebraic equation for the dominant eigenvalue is the first important  result of this letter. 

We mention a few classes of systems which can be solved exactly. 
As a  first example, we consider systems without ``persistence", i.e., $q_{\mathbf{\sigma}}\equiv 1$, leading to:
\begin{equation}
    \lambda=\sum_{\mathbf{\sigma}}p_{\mathbf{\sigma}}e^{-\beta \epsilon_{\mathbf{\sigma}}}.\label{fjc}
\end{equation}
The 3-d freely-jointed chain 
is obtained with the following identifications: $\mathbf{\sigma}$ represents the space angle $\mathbf{\Omega}$, specifying the orientation of each subunit (of fixed length $b$). Since this orientation is random, one has $p_\mathbf{\Omega}=1/\left(4 \pi \right)$, while the sum becomes an integral over the space angle, $\sum_{\mathbf{\sigma}}\rightarrow \int_0^{2\pi}d\phi \int_0^{\pi} d\theta \sin \theta  $. Furthermore $\epsilon_\mathbf{\Omega}=Fb\cos \theta $ represents the effect of an external field, with $\theta$ the angle between the monomer and this field.  Eq.~(\ref{fjc}) gives the result:
\begin{equation}
\lambda=\frac{\sinh(bF \beta)}{bF \beta}.\label{FJXD}
\end{equation}
By combination with Eq.~(\ref{fveeq}) and Eq.~(\ref{PF}), one recovers the aforementioned result $x=X/(Nb)= \mathcal{L}(b F \beta)$. 

The above analysis can be reproduced for state-independent persistence, i.e., $q \leq1$ but independent of $\mathbf{\sigma}=\mathbf{\Omega}$, by starting from Eq.~(\ref{ceq})  rather than from Eq.~(\ref{fjc}). The integral determining the eigenvalue $\lambda$,
\begin{equation}
\int^{2\pi}_0 d\phi \;\int^{\pi}_0 d\theta\,\frac{\sin \theta}{4\pi} \frac{q}{e^{bF \beta \cos \theta}\lambda-(1-q)}=1,
\end{equation}
can still be solved:
\begin{equation}
     \lambda=\frac{\left(e^{{2bF\beta}/{q}}-1\right)(1-q)}{e^{bF\beta\left({2}/{q}-1\right)}-e^{bF\beta}}.
     \end{equation}
The fractional extension $x=\mathcal{L}_q(b F \beta)$ is described by a generalized Langevin function, see also Fig.~\ref{langevinFig}:
     \begin{equation}
     \mathcal{L}_q(y)=\frac{\left(2-q\right)e^{{2y}/{q}}\left(e^{{2y}}-1\right)+q\left(e^{2y}-e^{{4y}/{q}}\right)}{q \left(e^{{2y}/{q}}-1\right)\left(e^{2y}-e^{{2y}/{q}}\right)}.\label{GenLan}
\end{equation}
The freely-jointed chain is retrieved in the limit $q\rightarrow 1$, with $ \mathcal{L}_q$ converging to $\mathcal{L}$. In the limit of strong persistence, $q \rightarrow 0$, $ \mathcal{L}_q$  converges to the sign function $ \mathcal{L}_0(y)=\textrm{sgn}(y)$.
For small forces, the polymer behaves as a harmonic spring, with spring constant $\kappa_q=F/ X$ given by:
\begin{equation}
     \kappa_q=\frac{q}{(2-q)}\kappa_1\;\;\;\text{with}\;\;\; \kappa_1=\frac{3}{Nb^2\beta}.
\end{equation}
$\kappa_1$ is the spring constant of the 3-d freely-jointed chain. Note the weakening of the spring for increasing persistence, corresponding to a decreasing value of $q$. These predictions have been verified using Langevin simulations, cf.~Fig.~\ref{langevinFig}.
Other applications, e.g., 1-d polymers and systems with only two $\mathbf{\sigma}$-states, are presented in the supplemental material. 

Depending on the interpretation of the model, the above transfer matrix describes  discrete steps taking place in space (for example in an Ising spin or polymer chain problem), in time (for a Markov chain)
or in another possibly more abstract coordinate (for example an angle coordinate or a higher dimensional vectorial coordinate).  To simplify further the eigenvalue equation, we  focus on a ``hydrodynamic" limit.
For simplicity, we present it as taking place in the context of a single scalar spatial variable. We associate an elementary spatial displacement of length $dl$ (corresponding to the bond length $b$ in the above polymer problem) to each discrete step. This length will be small compared to the average length of ``straight" segments, belonging to  a given state  $\mathbf{\sigma}$, $\bar{l}_{\mathbf{\sigma}}$, i.e., $dl/\bar{l}_{\mathbf{\sigma}}\rightarrow 0$. Meanwhile, the total length $L=Ndl$ will become large compared to the typical lengths of the problem, $\bar{l}_{\mathbf{\sigma}}/L\rightarrow 0$. In this limit, the ``jump probabilities" $q_{\mathbf{\sigma}}$ are replaced by transition probabilities per unit length $k_{\mathbf{\sigma}}=1/\bar{l}_{\mathbf{\sigma}}$:
\begin{eqnarray}
 q_{\mathbf{\sigma}}=k_{\mathbf{\sigma}} dl,\label{ct1}
  \end{eqnarray}
resulting in ``straight" segments with lengths that are exponentially distributed.
 The matrix $\mathbf{A}$ is replaced by a transition matrix $\mathbf{K}$:
\begin{equation}
 \mathbf{A}_{\mathbf{\sigma}',\mathbf{\sigma}}=\mathbf{1}+ \mathbf{K}_{\mathbf{\sigma}',\mathbf{\sigma}}    dl, \;\;\;\;
  \mathbf{K}_{\mathbf{\sigma}',\mathbf{\sigma}}=k_{\mathbf{\sigma}}\left(p_{\mathbf{\sigma'}}-\delta_{\mathbf{\sigma},\mathbf{\sigma'}}\right).\label{ct3}
  \end{equation}
Consistent with this limit, the  energy $\epsilon_{\mathbf{\sigma}}$ and the eigenvalue $\lambda$ converge as follows to $0$ and $1$, respectively:
\begin{equation}
\epsilon_{\mathbf{\sigma}}= F_{\mathbf{\sigma}} dl, \;\;\;\;\lambda=1+ \mu dl,
  \end{equation}
 with $ F_{\mathbf{\sigma}}$ representing the external ``force" or ``energy density" when the system is in the state $\mathbf{\sigma}$. 
 The eigenvalue equation Eq.~(\ref{ceq}) thus reduces to the  following algebraic relation for $\mu$:
\begin{equation}
\sum_{{\mathbf{\sigma}}} \frac{p_{\mathbf{\sigma}}k_{\mathbf{\sigma}}}{\mu+\beta F_{\mathbf{\sigma}}+k_{\mathbf{\sigma}}}=1.\label{cceq}
\end{equation}
We note that the dependence on the external field contribution $F_{\mathbf{\sigma}}$  is no longer via an exponential function. 
The corresponding free energy, cf.~Eq.~(\ref{PF}), becomes:
\begin{equation}\label{Gceq}
    \mathcal{G}\sim-\frac{L\mu}{\beta},
\end{equation}
with a simple proportionality to the  eigenvalue $\mu$. The extensivity in $N$ is replaced by an extensivity in $L$. Eq.~(\ref{Gceq}) is an equality to dominant order in $L$.

\begin{figure}
\begin{subfigure}{0.5\textwidth}
    
    \includegraphics[scale=0.55]{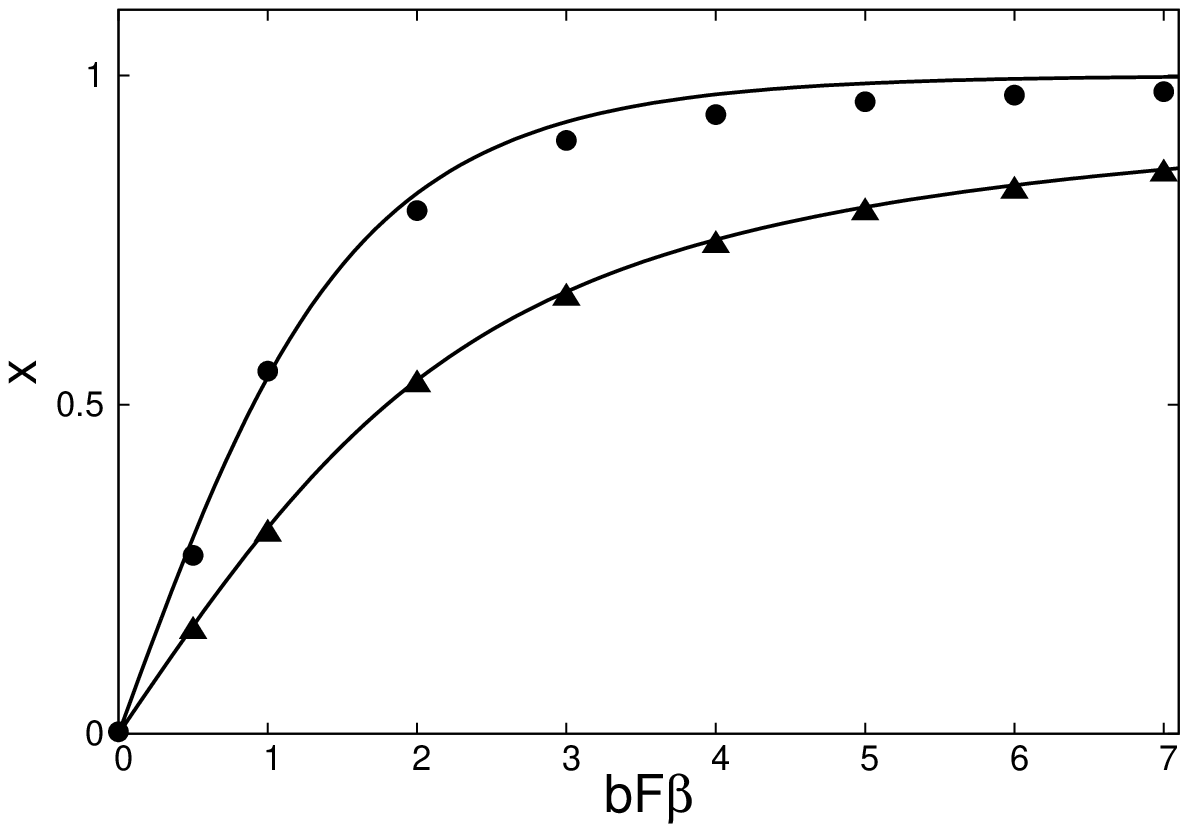}
    \caption{}
    \label{langevinFig}
\end{subfigure}\\%
\begin{subfigure}{0.5\textwidth}
    \centering
    \includegraphics[scale=0.55]{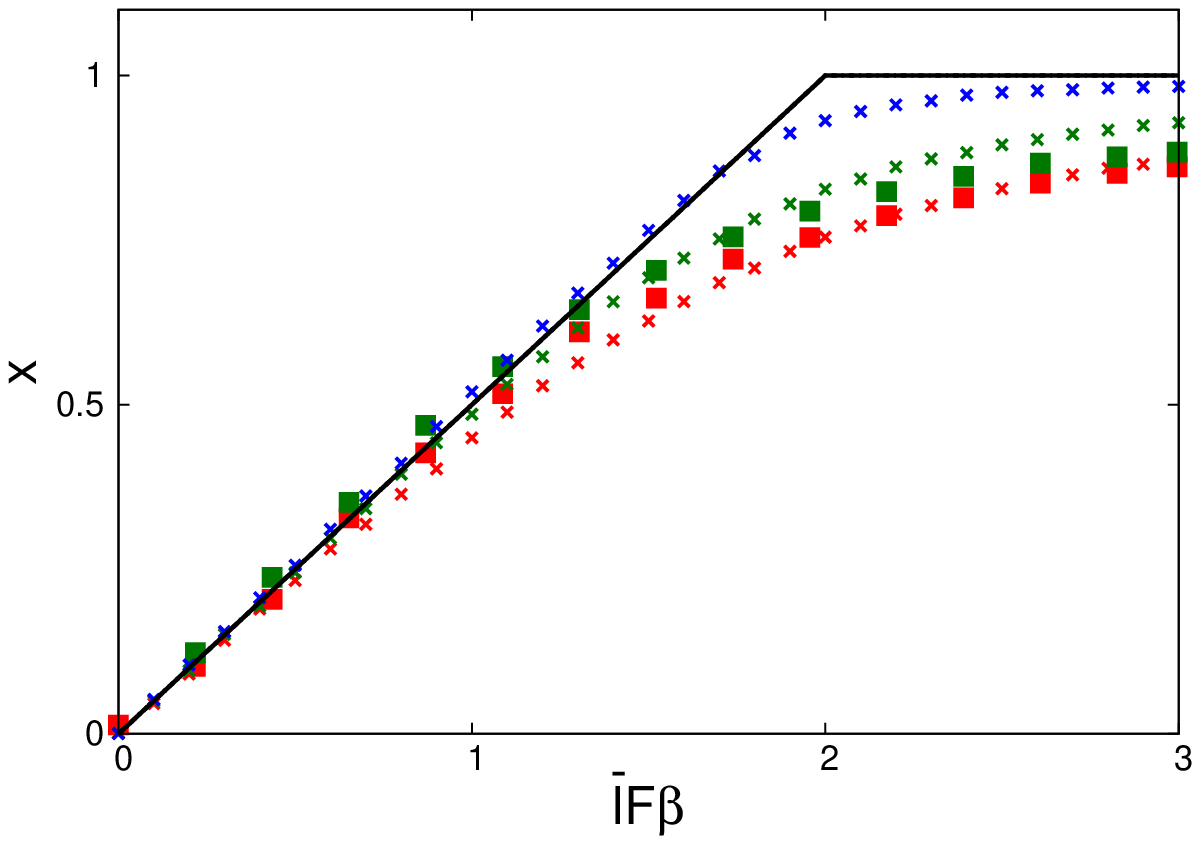}
    \caption{}
    \label{fig1}
\end{subfigure}
\caption{Extension $x$ versus force $F$ for 
    (a) $d=3$ freely-jointed chain $q=1$ and its modified version with persistence $q=0.7$:  theoretical result  Eq.~(\ref{GenLan}) (full line) versus Langevin simulation for $N=100$ (triangles $q=1$ and dots $q=0.7$).
    (b) random walk polymer model with persistence and appropriate transverse field reproducing Eq.~(\ref{pres}).  $L=10\, dl$ (red), $L=20\, dl$ (green) and $L=100\, dl$ (blue), with $\bar{l}/dl=10$ (Monte Carlo, crosses) and $\bar{l}/dl=5$ (Langevin, squares).}
\end{figure}

We now turn to the search for a ``perfect spring", defined as a system for which the free energy Eq.~(\ref{Gceq}) is exactly quadratic in the external field amplitude. More precisely, we  introduce the overall force  $F$ via the specification $F_{\mathbf{\sigma}}=a_{\mathbf{\sigma}}F$, with $a_{\mathbf{\sigma}}$ an $F$-independent amplitude, and require that 
\begin{equation}\mu=l(\beta F)^2\label{muhareq},\end{equation}
with  $l$ an $F$-independent reference length scale.  
The question thus reduces to finding probability distributions $p_{\mathbf{\sigma}}$ such that Eq.~(\ref{cceq}) holds under this constraint. By Taylor expansion in $F$, and under the assumption that $k_{\mathbf{\sigma}}={1}/{\bar{l}}$ is independent of the state $\mathbf{\sigma}$, one finds the following explicit expression for the moment generating function associated with $p_{\mathbf{\sigma}}$ (see supplemental materials):
\begin{equation}
\sum_{\mathbf{\sigma}} p_{\mathbf{\sigma}}e^{-\lambda  a_{\mathbf{\sigma}}}=\sqrt{\frac{\bar{l}}{l}}\; \frac{I_1\left(2 \sqrt{\frac{l}{\bar{l}} }\lambda\right)}{\lambda}.\label{lpsp}
\end{equation}
$I_1$ is the modified Bessel function.
This general criterion for harmonicity is our second major result.
The probability distribution $p_{\mathbf{\sigma}}$ can  be found from it, depending on the topology of the phase space and the form of $a_{\mathbf{\sigma}}$, by an inverse (integral) transform. Eq.~(\ref{lpsp}) obviously has no solution for a finite state space of $\mathbf{\sigma}$.
In particular, two-state models (corresponding for example to a polymer model in $d=1$) can not be turned into fully harmonic springs.

Combining Eq.~(\ref{fveeq}), Eq.~(\ref{Gceq}) and Eq.~(\ref{muhareq}),  one finds that the corresponding stretching fraction $x$ is given by:
\begin{equation}
x=X/L=-\frac{1}{L}\frac{d}{d F}\mathcal{G}=2{lF\beta },\;\;\;\;x\leq1.\label{xpf}\\
\end{equation}
The extension is thus exactly linear in $F$. The above result is only valid up to $x=1$, i.e., until the polymer is fully stretched. The reason for this limitation is that the Taylor expansion of Eq.~(\ref{cceq}) has a radius of convergence given by $F_c=1/(2l\beta)$. 
 For larger values of $F$, $x$ stays put at its maximal value $x=1$, hence the system undergoes a second order phase transition at $F=F_c$ (discontinuous second derivative of $\mathcal{G}$). 
 The spring constant $ \kappa$, corresponding to the harmonic law Eq.~(\ref{xpf}), is given by
  $\kappa={1}/(2Ll \beta)$. 
The above result has been derived in the  limit $dl/\bar{l} \rightarrow 0$. In the supplemental material, we evaluate the first order correction in $dl$ and conclude that the  harmonic behavior prevails, but with a modified spring constant:
\begin{equation}
     \kappa=\frac{1}{2Ll\beta}\left(1-2\frac{dl}{\bar{l}}\right).
\end{equation}

As a concrete application of the above harmonicity criterion, we return to the polymer problem in d-dimensional Euclidean space, with the identification of $\mathbf{\sigma}$ as a d-dimensional spatial angle $\mathbf{\Omega}$. Identifying $b$  with $dl$, one finds from  $\epsilon_\mathbf{\Omega}=Fb \cos \theta$ that  $a_\mathbf{\Omega}=\cos\theta$. 
If we furthermore assume that $p_\mathbf{\Omega}$ only depends on $\theta$, the integral Eq.~(\ref{lpsp}) can be solved  by inverting the integral transform.
Anticipating that $l=\bar{l}/4$ and comparing the harmonicity criterion, Eq.~(\ref{lpsp}), with the following integral representation of the Bessel function \cite{weisstein2002crc}:
\begin{equation}
I_1(\lambda)=\frac{\lambda}{\pi}\int_0^\pi d \theta \,e^{-\lambda \cos \theta} \sin^2 \theta,\label{Bessel}
\end{equation}
we conclude (remembering that the Jacobian of the $d$-sphere features  the factor  $\sin^{d-2}\theta$):
\begin{eqnarray} p_\mathbf{\Omega}&=&\mathcal{N}_d\,{\sin^{4-d} \theta}.\label{pres}
\end{eqnarray}
 ($\mathcal{N}_d$ is a normalisation constant.)
 This is our third major result. We conclude that a polymer with persistence, consisting of exponentially distributed straight segments, is perfectly harmonic until full stretching in $d=4$. In $d=3$ one needs an additional field, orthogonal to the stretching direction, which induces a biased distribution $p_\mathbf{\Omega}\sim {\sin \theta}$. Such a field can be realised by application of an electromagnetic force, which is often used in the experimental stretching of polymers. We have verified the latter prediction via Monte Carlo and Langevin simulations. The numerical results are in perfect agreement with the theory, cf.~Fig.~\ref{fig1} and supplemental material for more details. These conclusions are of course not restricted to polymer models.

In conclusion, we have introduced  ``Kubo-Anderson" transfer matrices, for which the largest eigenvalue can be obtained from the simple, explicit algebraic equation, Eq.~(\ref{ceq}). The latter simplifies upon taking a continuous limit to  Eq.~(\ref{cceq}).  By assuming a uniform transition rate, a simple, explicit criterion for harmonicity results, cf.~Eq.~(\ref{lpsp}). As a concrete example, we show that a polymer chain with persistence can behave, until fully stretched, as a perfect harmonic spring. Although our model is rather theoretical, and perhaps not experimentally feasible, its discovery can serve as the starting point for the construction of complex polymer systems with enhanced harmonicity. These results can be easily mapped on other systems. For example, one could construct an ising-like magnet with perfectly linear susceptibility.

%

\pagebreak
\newpage
\widetext\newpage
\begin{center}
\textbf{\large Supplemental Materials}
\end{center}
\setcounter{equation}{0}
\setcounter{figure}{0}
\setcounter{table}{0}
\setcounter{page}{1}
\makeatletter
\renewcommand{\theequation}{S\arabic{equation}}
\renewcommand{\thefigure}{S\arabic{figure}}
\renewcommand{\bibnumfmt}[1]{[S#1]}
\renewcommand{\citenumfont}[1]{S#1}

 \affiliation{Hasselt University, B-3590 Diepenbeek, Belgium.}
 
\date{\today}

\maketitle

\section{Two state model}
 For a two-state system,  $\sigma=\pm$, Eq.~(\ref{ceq}) for the eigenvalue $\lambda$ reduces to:
\begin{equation}
    \frac{p_{+}q_{+}}{e^{\beta \epsilon_{+}}\lambda-(1-q_{+})}+\frac{p_{-}q_{-}}{e^{\beta \epsilon_{-}}\lambda-(1-q_{-})}=1.
\end{equation}
The relevant solution of this quadratic equation in $\lambda$ is given by:
\begin{equation}
   \lambda=\frac{\left( e^{-\beta\epsilon_-}\left(1-p_+q_-\right)+e^{-\beta\epsilon_+}\left(1-p_-q_+\right)\right)}{ 2} \left(1+\sqrt{1-\frac{4e^{-\beta\left(\epsilon_++\epsilon_-\right)}\left(1-p_+q_--p_-q_+\right)}{\left( e^{-\beta\epsilon_+}\left(p_+q_--1\right)+e^{-\beta\epsilon_-}\left(p_-q_+-1\right)\right)^2}}\right).
\end{equation}
The free energy follows by combination with Eq.~(\ref{PF}), and the force versus extension relation is then obtained from Eq.~(\ref{fveeq}).

As an example, we consider an unbiased random walker with persistence. Setting
$p_+=p_-=1/2$, $q_+=q_-=q$ and $\epsilon_+=-\epsilon_-=Fb$ in the above expression, one finds:
\begin{equation}
    \lambda=\left(1-\frac{q}{2}\right)\cosh\left(bF \beta\right)\left(1+\sqrt{1-\frac{(1-q)}{\left(1-\frac{q}{2}\right)^2\cosh^2(bF \beta)}}\right).
\end{equation}
The resulting fractional extension reads:
\begin{equation}x=\frac{\left(1-\frac{q}{2}\right)\sinh(bF \beta)}{\sqrt{\left(1-\frac{q}{2}\right)^2\cosh(bF \beta)^2+\left(1-q\right)}}.\label{xpers}\end{equation}
For small forces, one can associate a spring constant to this system equal to
\begin{equation}
    \kappa_q=   \frac{q}{ 2-q}\kappa_1,\;\;\;\;\ \kappa_1=\frac{1}{N b^2\beta }.
\end{equation}
In the absence of persistence, i.e., in the limit $q\rightarrow 1$, one reproduces the $d=1$ random walk result $x=\tanh \left(bF\beta\right) $, while $x=\textrm{sgn}(bF\beta)$
in the "complementary" limit $q\rightarrow 0$.

\section{Temperature and Energy dependence}
To illustrate the dependence of the parameters $q_{\mathbf{\sigma}}$ and $p_{\mathbf{\sigma}}$ on energy and temperature, we consider the case of constant $q_{\mathbf{\sigma}}=q$ and $p_{\mathbf{\sigma}}=p$, independent of ${\mathbf{\sigma}} $. Introducing the interaction energies $E^{+}$, upon staying in the same state, and $E^{-}$, upon switching states, both again assumed to be independent of the state (and $E^{+}\leq E^{-}$), one readily finds:
\begin{eqnarray}
q=1-e^{-\beta E^{+}}+e^{-\beta E^{-}},\;\;\;
p=\frac{e^{-\beta E^{-}}}{1-e^{-\beta E^{+}}+e^{-\beta E^{-}}}.
\end{eqnarray}

\section{Freely-jointed 4d chain}
Applying the eigenvalue Eq.~(\ref{ceq}) to the four dimensional freely jointed chain, one finds
\begin{equation}
\int_0^{2\pi} d\phi'\int_0^{\pi} d\phi\int^{\pi}_0d\theta\,\frac{\sin^2 \theta\sin \phi}{2\pi^2} \frac{q}{e^{\beta b \cos \theta}\lambda-(1-q)}=1.
\end{equation}
For the case without persistence, $q=1$, the integral simplifies to
\begin{equation}
\lambda=\int_0^{2\pi} d\phi'\int_0^{\pi} d\phi\int^{\pi}_0d\theta\,\frac{\sin^2 \theta\sin \phi}{2\pi^2} e^{-\beta b \cos \theta}=\frac{2I_1(bF \beta)}{bF \beta},
\end{equation}
and the fractional extension becomes
\begin{equation}
x=\frac{bF \beta\left(I_0(bF \beta)+I_2(bF \beta)\right)}{2 I_1(bF \beta)}-\frac{1}{bF \beta}.
\end{equation}

By considering the continuous limit,  using Eq.~(\ref{cceq}), as departure point, one finds for the eigenvalue $\mu$:
\begin{equation}
\int_0^{2\pi} d\phi'\int_0^{\pi} d\phi\int^{\pi}_0d\theta\,\frac{\sin^2 \theta\sin \phi}{2\pi^2} \frac{k}{\mu+\beta F \cos \theta+k}=1.
\end{equation}
Using the following definite integral:
\begin{equation}
\int_0^\pi  d\theta \frac{\sin^2 \theta}{(a+ b \cos \theta)}=\frac{\pi \left(a-\sqrt{a^2-b^2}\right)}{b^2},
\end{equation}
we recover the result of the main text:
\begin{equation}
\mu=\frac{\left(\beta F\right)^2}{4k}.
\end{equation}

\section{Finite $dl$ correction}
To show the robustness of the perfectly harmonic regime, we evaluate the effect of a finite value of $dl/\bar{l}$, by calculating a first order correction. We expand Eq.~(\ref{ceq}) for  $\lambda$, as in the main text. $\lambda$ is expanded one order further in $dl$:
\begin{equation}
    \lambda=1+\mu dl+\nu dl^2.
\end{equation}
Making a Taylor expansion up to second order in $dl$, one finds:
\begin{eqnarray}
\sum_{\mathbf{\sigma}} \frac{k_{\mathbf{\sigma}} p_{\mathbf{\sigma}} }{\mu+\beta F_{\mathbf{\sigma}}+k_{\mathbf{\sigma}}}&=&1,\\\sum_{\mathbf{\sigma}}
\frac{k_{\mathbf{\sigma}} p_{\mathbf{\sigma}}\left(2\nu +2\beta\mu F_{\mathbf{\sigma}}+\beta^2 F^2_{\mathbf{\sigma}}\right)}{\left(k_{\mathbf{\sigma}}+\mu+\beta F_{\mathbf{\sigma}} \right)^2}&=&0.\label{eqnu}
\end{eqnarray}
The first equation determines the value of $\mu$, For the perfect spring model, discussed in the main text, this gives:
\begin{equation}
    \mu={{l}(\beta F)^2}=\frac{\alpha (\beta F)^2}{4k},\label{mudef}
\end{equation}
under the assumption that $l$ is inversely proportional to $k$, $l=\alpha/k$.
If we assume that $k_{\mathbf{\sigma}}=k$, and that $k, p_{\mathbf{\sigma}}$ and $F_{\mathbf{\sigma}}$ are independent of $\beta$, one can show, by taking the derivatives of the first equation to $k$ and $\beta$,
\begin{equation}
\sum_{\mathbf{\sigma}} \frac{k p_{\mathbf{\sigma}} }{\left(\mu+\beta F_{\mathbf{\sigma}}+k\right)^2}=\frac{1}{k\left(1+\frac{d}{dk}\mu\right)},
\end{equation}
\begin{equation}
\sum_{\mathbf{\sigma}} \frac{k p_{\mathbf{\sigma}} F_{\mathbf{\sigma}} }{\left(\mu+\beta F_{\mathbf{\sigma}}+k\right)^2}=-\sum_{\mathbf{\sigma}} \frac{k p_{\mathbf{\sigma}}  }{\left(\mu+\beta F_{\mathbf{\sigma}}+k\right)^2}\frac{d}{d\beta} \mu.
\end{equation}
This allows to rewrite Eq.~(\ref{eqnu}), determining the value of $\nu$ as
\begin{equation}
2\nu \sum_{\mathbf{\sigma}} \frac{k p_{\mathbf{\sigma}} }{\left(\mu+\beta F_{\mathbf{\sigma}}+k\right)^2}+k-\sum_{\mathbf{\sigma}} \frac{k p_{\mathbf{\sigma}} \left(\left(\mu+k\right)^2+2\beta k F_{\mathbf{\sigma}}\right)}{\left(\mu+\beta F_{\mathbf{\sigma}}+k\right)^2}=0,
\end{equation}
giving
\begin{equation}
\nu=\frac{\left(\mu+k\right)^2-2\beta k \frac{d}{d\beta}\mu-k^2-k^2\frac{d}{dk}\mu}{2}.
\end{equation}
From Eq.~(\ref{mudef}), one deduces
\begin{equation}
\nu=\frac{\left(\beta F\right)^4}{32k^2}+\frac{\left(\beta F\right)^2}{8},
\end{equation}
and therefore,
\begin{equation}
\ln\lambda\approx\ln\left(1+\mu dl+\nu dl^2\right)\approx\mu dl+\left (\nu-\frac{\mu^2}{2}\right)dl^2=\frac{\alpha (\beta F)^2}{k} dl+\frac{kl(\beta F)^2}{2}dl^2,
\end{equation}
or
\begin{equation}
    \kappa=\frac{1}{2Ll\beta}\left(1-2kdl\right)
\end{equation}

\section{Moments of perfect spring}
The set of equations for the moments of the probability distribution for the harmonicity criterion are found by expanding:
\begin{equation}
    \sum_{\boldsymbol{\sigma}}\frac{p_{\boldsymbol{\sigma}}k_{\boldsymbol{\sigma}}}{k_{\boldsymbol{\sigma}}+a_{\boldsymbol{\sigma}}\beta F+l(\beta F)^2}=1
\end{equation}
in a Taylor series of $F$ followed by a binomial expansion:
\begin{eqnarray}
&&\sum_{\boldsymbol{\sigma},n}p_{\boldsymbol{\sigma}}(-1)^n\left(\frac{a_{\boldsymbol{\sigma}}\beta F+l(\beta F)^2}{k_{\boldsymbol{\sigma}}}\right)^n=\sum_{\boldsymbol{\sigma},n}p_{\boldsymbol{\sigma}}\left(-\frac{1}{k_{\mathbf{\sigma}}}\right)^n\sum_{m=0}^n\left(\begin{array}{c}
    n  \\
     m
\end{array}\right)a_{\boldsymbol{\sigma}}^ml^{n-m}(\beta F)^{2n-m}=1.
\end{eqnarray}
By regrouping terms with the same exponent of $\beta F$
we derive the following set of conditions:
\begin{equation}
    \sum_{m=0}^{n}(-1)^{m}\left(\begin{array}{c}
        n+m\\
        2m
    \end{array}\right)\left\langle\frac{a^{2m}_{\mathbf{\sigma}}}{l^{m}k_{\mathbf{\sigma}}^{n+m}}\right\rangle =0,\;\;\;\forall n\in\mathbb{N},\;\;\;n\neq 0,\label{mmaq}
\end{equation}
with the brackets referring to an average with respect to $p_{\mathbf{\sigma}}$.
A situation of particular interest is the case where $k_{\mathbf{\sigma}}={1}/{\bar{l}}$ is independent of the state $\mathbf{\sigma}$. $\bar{l}$ is the (now $\mathbf{\sigma}$-independent) average persistence length. This set of equations can be solved using the following mathematical identity:
\begin{equation}
\sum_{m=0}^n(-1)^m\frac{(n+m)!}{(n-m)!m!(m+1)!}=0,
\end{equation}
which immediately implies
\begin{equation}\label{maeq}
    \left\langle{a^{2n}_{\mathbf{\sigma}}}\right\rangle=\left(\frac{l}{\bar{l}}\right)^n\frac{(2n)!}{n!(n+1)!},
\end{equation}
while all odd moments $\langle a^{2n+1}_{\mathbf{\sigma}}\rangle$ are all zero. The associated moment generating function is given by
\begin{eqnarray}
\left\langle e^{-\lambda a_{\mathbf{\sigma}}}\right\rangle&=&\sum_{k=0}^{\infty} \frac{\left\langle a^{k}\right\rangle\left(-\lambda\right)^k}{k!}=\sum_{k=0}^{\infty}\left(\frac{l}{\bar{l}}\right)^k \frac{\lambda^{2k}}{k!(k+1)!}.
\end{eqnarray}
Identification with the following expansion of the modified Bessel function:
\begin{eqnarray}
I_1(\lambda)=\sum_{k=0}^{\infty}\frac{1}{(k+1)!k!}\left(\frac{\lambda}{2}\right)^{2k+1},
\end{eqnarray}
leads to the result Eq.~(\ref{lpsp}) of the main text.

\section{Monte Carlo simulations}
We first generate an array of length $N$ with the directions of the individual monomers chosen at random. At each subsequent step of the Monte Carlo simulation, we randomly choose an element $i$ from the array, and update its state according to the "familiar" Monte Carlo rules
\begin{itemize}
    \item If $i=1$ or $N$, the site will be in the same state as the neighbouring site with probability
    \begin{equation}
        p_{\theta}=\frac{e^{Fb\cos(\theta)}(1-k)}{e^{Fb\cos(\theta)}(1-k)+k\int^{\pi}_0d\theta' \left(\sin \theta'\right)^2e^{Fb\cos(\theta')}}
    \end{equation}
    and in a state drawn from the distribution \begin{equation}\left(\sin \theta'\right)^2\exp{\left(Fb\cos(\theta')\right)}\label{pnew}\end{equation}
    with probability $1-p_{\theta}$.
    \item If the site on the left of $i$ is in the same state, $\theta$, as the site on the right of $i$, $i$ will also be set to this state with probability
    \begin{equation}
        p_{\theta}=\frac{e^{Fb\cos(\theta)}(1-k)^2}{e^{Fb\cos(\theta)}(1-k)^2+k^2\int^{\pi}_0d\theta' \left(\sin \theta'\right)^2e^{Fb\cos(\theta')}}
    \end{equation}
    \item If the state on the left and on the right of $i$ differ, the probability to be set in the state $\theta_{i-1}$ or $\theta_{i+1}$ is equal to
    \begin{eqnarray}
        p_{\theta_{i-1}}&=&\frac{e^{Fb\cos(\theta_{i-1})}(1-k)}{\left(e^{Fb\cos(\theta_{i-1})}+e^{Fb\cos(\theta_{i+1})}\right)(1-k)+k\int^{\pi}_0d\theta' \left(\sin \theta'\right)^2e^{Fb\cos(\theta')}},\\
        p_{\theta_{i+1}}&=&\frac{e^{Fb\cos(\theta_{i+1})}(1-k)}{\left(e^{Fb\cos(\theta_{i-1})}+e^{Fb\cos(\theta_{i+1})}\right)(1-k)+k\int^{\pi}_0d\theta' \left(\sin \theta'\right)^2e^{Fb\cos(\theta')}},
    \end{eqnarray}
    and with probability $1-p_{\theta_{i-1}}-p_{\theta_{i+1}}$ from the probability distribution Eq.~(\ref{pnew}).
    
\end{itemize}
This procedure is repeated until convergence, typically for $10^{7}L/(kdl)^2$ steps.

\section{Langevin simulations}

We apply the Langevin formalism to construct a numerical scheme that achieves the most experimentally viable realisation of a perfect spring. We model the spring as a chain of $N+1$ beads with mass $m$ and position vector $\vec{x}_i$, with $i=0,1,\dots,N$. We take $\vec{x}_i\in {\rm I\!R}^4$, this reduces the complexity of the simulation and does not influence the resulting steady state behaviour. The beads interact through a potential $U(\vec{x}_0,\vec{x}_1,\dots,\vec{x}_N)=U(\{\vec{x}_i\})$. The Langevin equation describes the molecular dynamics of the beads while they are in contact with a viscous heat-bath with friction coefficient $\gamma$ and at temperature $T$. We assume overdamped dynamics where $m/\gamma \rightarrow 0$, the Langevin equation is then given by
\begin{equation}
\gamma\, \frac{d\vec{x}_i(t)}{dt} = - \vec{\nabla}U(\{\vec{x}_i\}) + \vec{\xi}_i(t). \label{LE}
\end{equation}
The second term on the right-hand side, $\vec{\xi}_i(t)$, represents the random collisions of the bath particles. This is a stochastic process (uncorrelated between beads) that is Gaussian distributed with zero mean and a correlation that is determined by the fluctuation-dissipation relation: $\langle \xi_{i,\alpha}(t)\xi_{j,\beta}(t')\rangle=2\gamma k_BT\delta(t-t')\delta_{i,j}\delta_{\alpha,\beta}$, with $k_B$ the Boltzmann constant and the second (Greek) indices indicates the specific components of the vector. 

Notice that this description introduces, in comparison with the theory, a new fundamental parameter $\gamma$, i.e. the friction coefficient. This parameter will, however, only affect the transient dynamics of the spring and not the steady state behaviour which is the relevant quantity for our discussion. 

The potential $U(\{\vec{x}_i\})$ should describe a perfect spring, it consists of two components: $U=U_h+U_p$. First, the distance between neighbouring beads should maintain a constant value $b$. We achieve this by linking them with a stiff harmonic spring with rest-length $b$ and a very large spring constant $k$. We have  
\begin{equation}
U_h(\{\vec{x}_i\}) = \frac{k}{2} \sum_{i=1}^{N+1} \Big( |\vec{r}_i| - b \Big)^2,
\end{equation}
with $\vec{r}_i=(\vec{x}_i-\vec{x}_{i-1})$ a bond-vector and $k \gg 1$. When $U_h$ is the only contribution to the total potential $U(\{\vec{x}_i\})$, the spring is called a freely-jointed chain. 

If, however, we want to describe a perfect spring, the following potential should be added  
\begin{equation}
U_p(\{\vec{x}_i\}) =- \frac{1}{\beta}\sum_{i=1}^{N}\,\ln\left[ \frac{2q}{\pi} + \frac{2(1-q)\delta(\hat{r}_i-\hat{r}_{i+1})}{\sin^2(\theta_i)}  \right],
\end{equation}
with $\beta=(k_BT)^{-1}$, $\theta=\cos^{-1}(\hat{r}_i\cdot\hat{r}_{i+1})$ and $q$ the chance of a bond to change direction. The Dirac delta function $\delta(\cdot)$ in this equation can not be implemented effectively in a numerical scheme, we therefore consider the following Gaussian which defines the Dirac delta function
\begin{equation}
\delta(x) = \lim_{\epsilon\rightarrow 0} \frac{1}{\sqrt{2\pi\epsilon^2\,}}\,e^{-x^2/2\epsilon^2}.
\end{equation}
This introduces a parameter $\epsilon$ which should be as small as possible. The ``perfect'' potential can thus be approximated by
\begin{equation}
U_p(\{\vec{x}_i\}) \approx - \frac{1}{\beta}\sum_{i=1}^{N}\,\ln\left[ \frac{2q}{\pi} + \frac{2(1-q)\exp(-(1-\cos(\theta_i))/\epsilon^2)}{\sqrt{2\pi\epsilon^2\,}\sin^2(\theta_i)}  \right], \label{perfU}
\end{equation}
with $\epsilon\ll 1$. We also used $(\hat{r}_i-\hat{r}_{i+1})^2=2-2\cos(\theta_i)$. 

The (first order) numerical integrator we used to find the time-evolution of the beads, i.e. $\vec{x}_i(t)$, resulting from equation \eqref{LE} is given by \cite{branka1999algorithms}
\begin{equation}
x_{i,\alpha}(t+\Delta t) = x_{i,\alpha}(t) - \frac{\Delta t}{\gamma} \, \frac{\partial\,U(\{\vec{x}_i\})}{\partial x_{i,\alpha}} + \eta(t).
\end{equation}
The stochastic term $\eta(t)$ is a Gaussian random variable with zero mean and a correlation given by $\langle \eta(t)\eta(t') \rangle=2k_BT\delta(t-t')\Delta t/\gamma$. Due to the large forces arising from small $\epsilon$ and large $k$, one should take the time-step $\Delta t$ appropriately small to avoid large numeric errors.

\begin{figure}
\center
\includegraphics[width=0.45\textwidth]{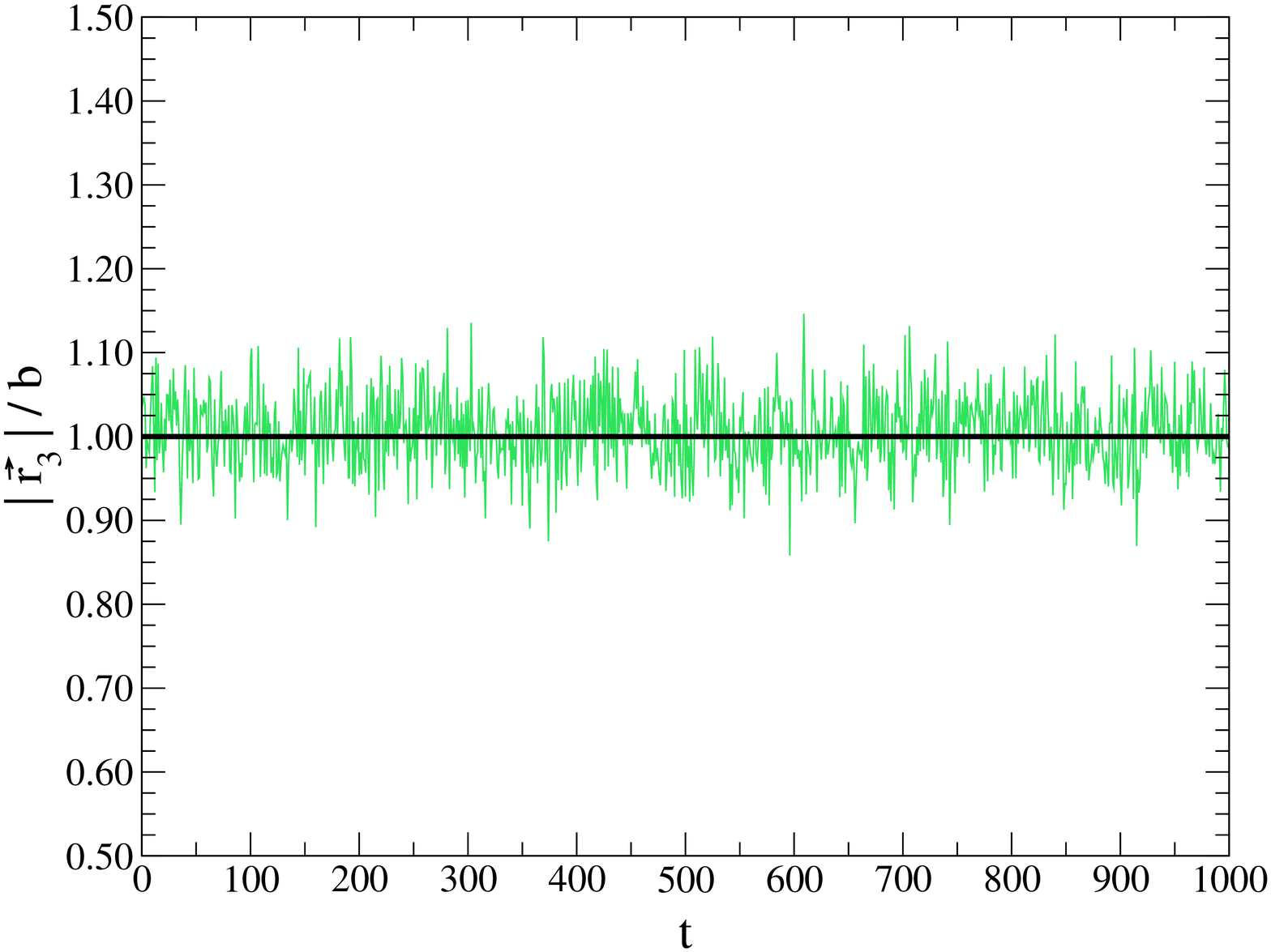}   
\includegraphics[width=0.45\textwidth]{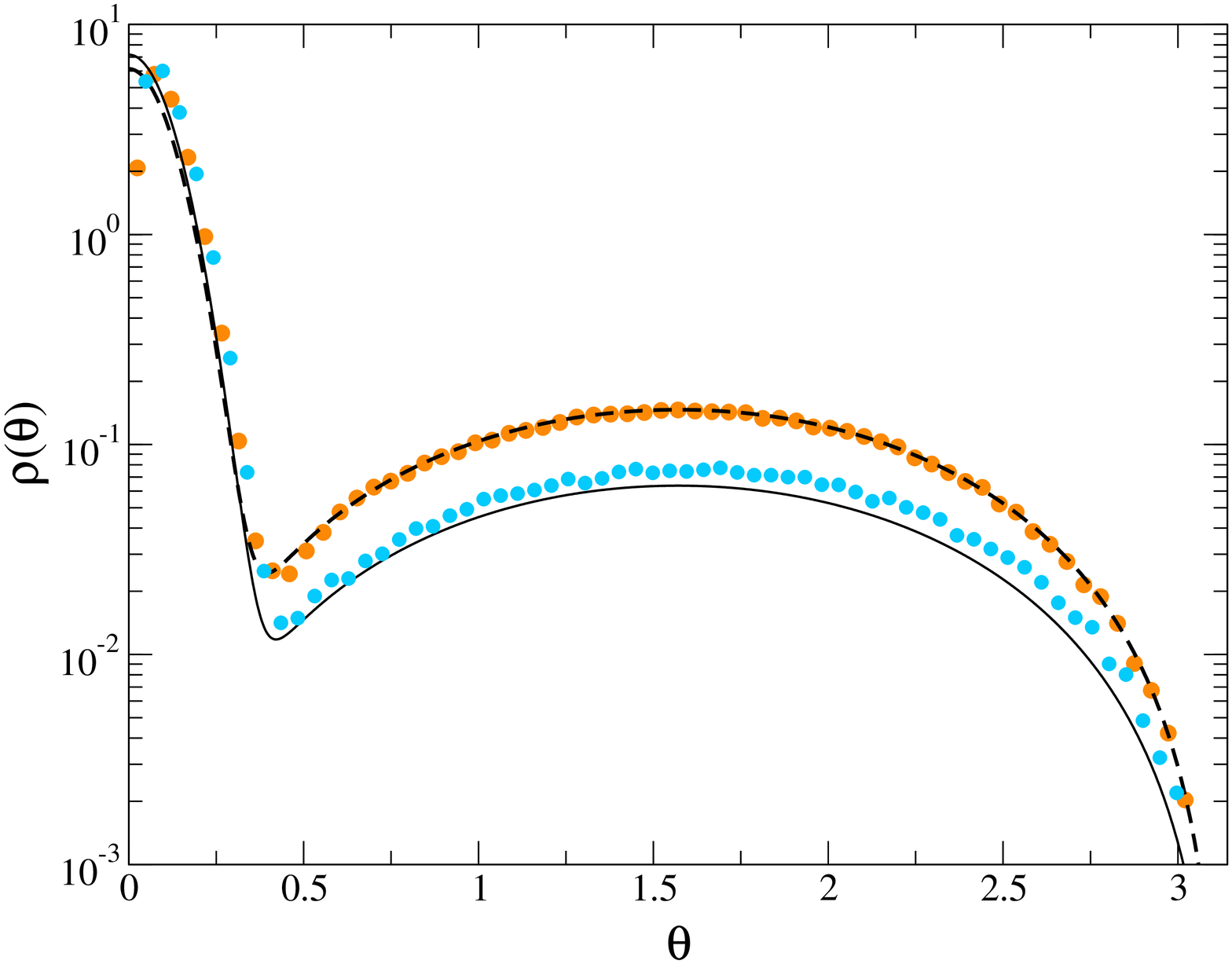}  
\caption{Both figures have the following parameters: $k_BT=\gamma=1$, $k=5\cdot 10^4$, $b=q=0.1$ and $\epsilon=0.1$.  {\em Left:} Simulation of the time-evolution of the relative bond-length of the third bond in a short chain of $N=5$ and $\Delta t=10^{-6}$. {\em Right:} Analytical probability distribution  for $q=0.1$ (full line) and $q=0.23$ (dashed line). Simulated probability distribution (from $5\cdot10^5$ samples) of the angle between two bonds (i.e. $N=2$), with $\Delta t=10^{-6}$ (orange dots) and $\Delta t=10^{-7}$ (blue dots).}  
\label{fig:test}     
\end{figure}

To examine the performance of our numerical scheme, we devised two tests. Firstly, we should check that the bond-lengths do not deviate greatly from the $b$-value. In the left graph of figure \ref{fig:test}, we plot the evolution of the relative bond-length, $|\vec{r}_i|/b$, of one bond as a function of time. When taking $k = 5\cdot 10^4$ and $\Delta t=10^{-6}$, we find that the relative bond-length never exceeds a deviation  of more than $10\%$ from its desired value of one. Secondly, one can investigate the distribution $\rho(\theta)$ of the angles between two consecutive bonds. From the ``perfect'' potential $U_p$ this distribution can be found using the Boltzmann factor: $\rho(\theta) = \mathcal{N} \exp(-\beta U_p)\sin^2(\theta)$, with $\mathcal{N}$ a normalisation factor. Note that here we take $U_p$ to be between only two bond-vectors, i.e. equation \eqref{perfU} without the summation. It is easy to see that $\mathcal{N}=1$, we therefore find
\begin{equation}
\rho(\theta) = \frac{2q}{\pi}\sin^2(\theta) + \frac{2(1-q)}{\sqrt{2\pi\epsilon^2\,}}\exp\left(-\frac{1-\cos(\theta)}{\epsilon^2}\right).
\end{equation}
This distribution is plotted (full line) in the right graph of figure \ref{fig:test}, for $\epsilon=b=q=0.1$. From this figure it is clear that the potential is sharply peaked around small angles, while it has a smaller - yet free - distribution for larger angles. Simulated distributions are also plotted in this figure for $\Delta t=10^{-6}$ and $\Delta t=10^{-7}$. One can see that a smaller time-step yields a better agreement with the analytical curve, while for larger $\Delta t$ the distribution of the bulk angles increases. This is due to the inaccuracy of the finite integration step, which is only completely resolved for $\Delta t \rightarrow 0$. To reach steady state in a reasonable computation time the time-step can not be too small, we therefore choose to work with $\Delta t=10^{-6}$. The deviation from the desired probability distribution can be fixed by assuming an effective $q$. From figure \ref{fig:test}, one can see that if we plot the distribution (dashed line) for $q=0.23$, it corresponds exactly with the less-accurate simulated distribution of $q=0.1$. We therefore accept $q=0.23$ as the effective $q$ of the simulated perfect spring.

In order to extract the force-extension relation of the perfect spring, we need to introduce two new forces in equation \eqref{LE}. We add a constant force $F$ in the $x$-direction to one of the end-beads (for example to $\vec{x}_N$), its integration step becomes
\begin{equation}
x_{N,\alpha}(t+\Delta t) = x_{N,\alpha}(t) - \frac{\Delta t}{\gamma} \, \frac{\partial\,U(\{\vec{x}_i\})}{\partial x_{N,\alpha}} + \frac{\Delta t}{\gamma}\, F\delta_{\alpha,x} + \eta(t).
\end{equation}
The other end-bead, $\vec{x}_0$, should remain fixed in the origin. We therefore assume that any force it experiences from the chain or the heat-bath will be counteracted by the origin. Or in other words, there is no net force acting upon this bead, so 
\begin{equation}
x_{0,\alpha}(t+\Delta t) = x_{0,\alpha}(t).
\end{equation}

The initial configuration of the spring is chosen random, only enforcing the constant $b$-length of the bonds. When the spring is allowed to evolve under the influence of the constant force, it will first go through a transient regime where it increases its extension. Thereafter, it arrives in a steady state where its average extension does not chance. The average extension $X(t)$ through time is computed as follows
\begin{equation}
X(t) = \left\langle \frac{L_x(t)}{L(t)} \right\rangle,
\end{equation}
where $L_x(t)=x_{N,x}(t)-x_{0,x}(t)$ and $L=|\vec{x}_N(t)-\vec{x}_0(t)|$. The averaging $\langle\cdot\rangle$ is done over different realisations of the spring. When the spring has reached steady state we write the average extension as $X=X(t\gg1)$.

\begin{figure}
\center
\includegraphics[width=0.45\textwidth]{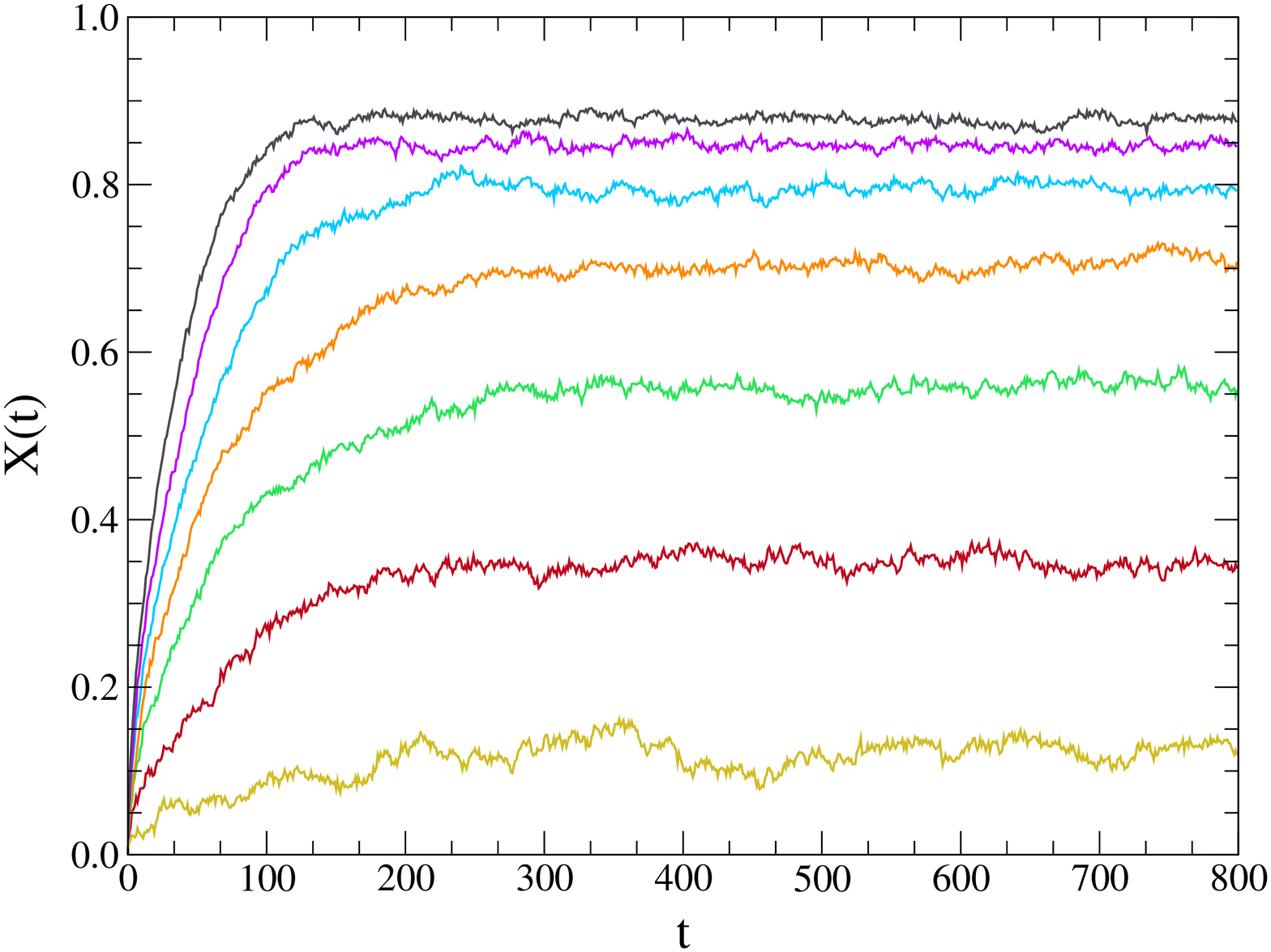}   
\includegraphics[width=0.45\textwidth]{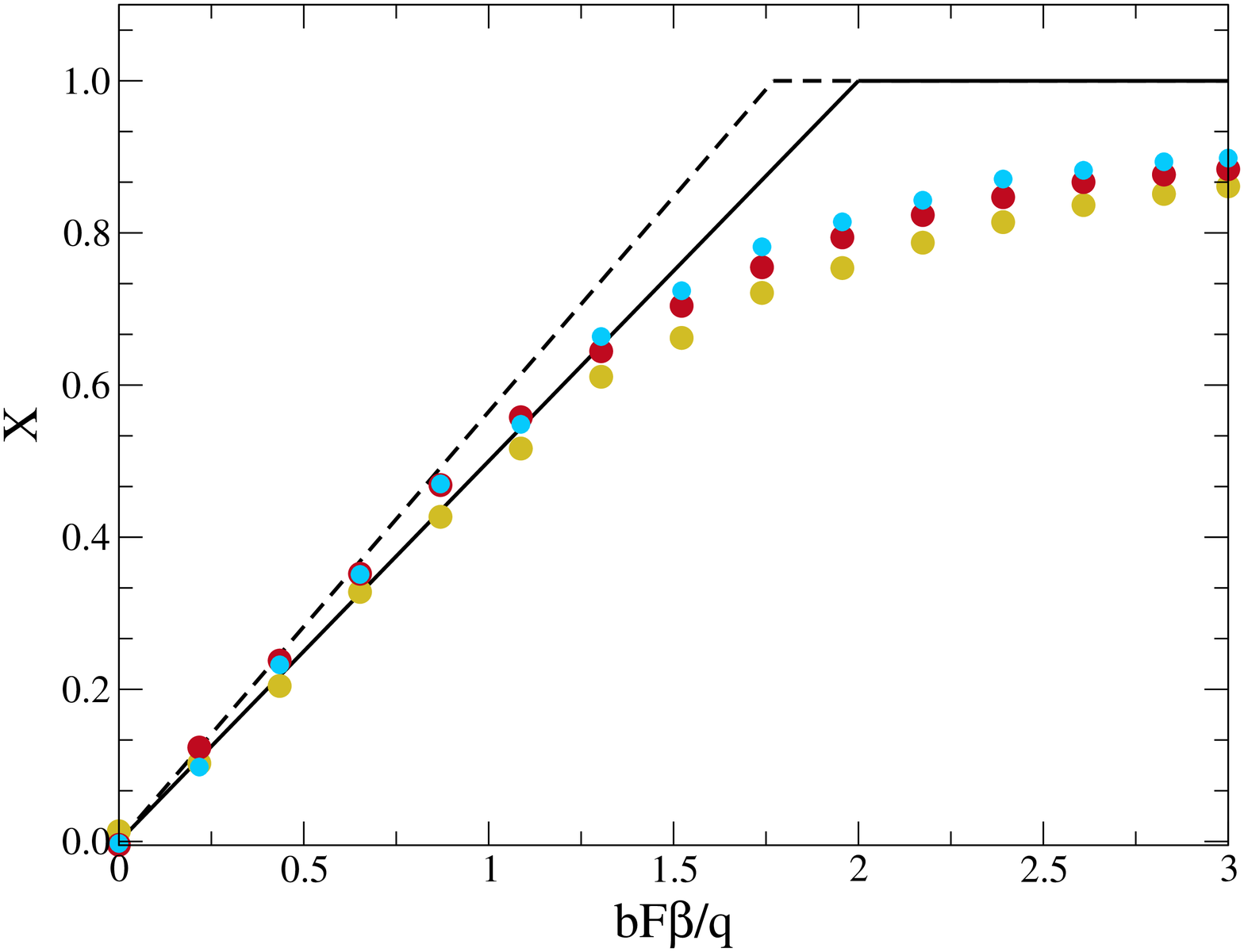}  
\caption{Both figures have the following parameters: $k_BT=\gamma=1$, $k=5\cdot 10^4$, $b=q=0.1$, $\epsilon=0.1$, $\Delta t=10^{-6}$ and effective $q=0.23$. {\em Left:} Simulation of the time-evolution of the extension $X(t)$ for $N=100$ and $F=0.5$ (yellow), $1.5$ (red), $2.5$ (green), $3.5$ (blue), $4.5$ (orange), $5.5$ (purple) and $6.5$ (grey). Averaged over $20$ histories. {\em Right:} Simulated force-extension of perfect springs with $N=50$ (yellow), $100$ (red) and $250$ (blue). Acquired from samples (in the order of $10^1-10^2$) of the steady state averaged extension. The full line is theoretical prediction (see main text) and the dashed line is the first order correction. }  
\label{fig:timeX}     
\end{figure}

From the time-plot (left graph in figure \ref{fig:timeX}) of the averaged extension, one can see that for all shown forces, $X(t)$ is able to reach steady state. When in steady state, we can obtain $X$ from the simulation. To acquire even more averaging, we also take different (uncorrelated) samples from the steady state regime. The right graph in figure \ref{fig:timeX} (see also main text) shows the force-extension for different values of $N$. For small forces they correspond well with the theoretical prediction (and even better with the first order correction). Deviations for larger forces are due to the smoothing of the delta function (i.e. $\epsilon \neq 0$) and the fact that $N$ is finite.     

\end{document}